\documentclass[twocolumn,twoside,slac]{revtex4}
\usepackage{graphicx}
\usepackage{fancyhdr}
\begin{document}

\title{Statistical Challenges in Modern Astronomy}

\author{E. D. Feigelson}
\affiliation{Department of Astronomy \& Astrophysics, Penn State
University, University Park PA 16802, USA}

\author{G. J. Babu}
\affiliation{Department of Statistics, Penn State University,
University Park PA 16802, USA}

\begin{abstract}

Despite centuries of close association, statistics and astronomy
are surprisingly distant today.  Most observational astronomical
research relies on an inadequate toolbox of methodological tools.
Yet the needs are substantial: astronomy encounters sophisticated
problems involving sampling theory, survival analysis,
multivariate classification and analysis, time series analysis,
wavelet analysis, spatial point processes, nonlinear regression,
bootstrap resampling and model selection.  We review the recent
resurgence of astrostatistical research, and outline new
challenges raised by the emerging Virtual Observatory.  Our essay
ends with a list of research challenges and infrastructure for
astrostatistics in the coming decade.

\end{abstract}

\maketitle \thispagestyle{fancy}

\section{The glorious history of astronomy and statistics}

Astronomy is perhaps the oldest observational science\footnote{The
historical relationship between astronomy and statistics is
described in references [15], [38] and elsewhere. Our {\it
Astrostatistics} monograph gives more detail and contemporary
examples of astrostatistical problems [3].}. The effort to
understand the mysterious luminous objects in the sky has been an
important element of human culture for at least $10^4$ years.
Quantitative measurements of celestial phenomena were carried out
by many ancient civilizations.  The classical Greeks were not
active observers but were unusually creative in the applications
of mathematical principles to astronomy.  The geometric models of
the Platonists with crystalline spheres spinning around the static
Earth were elaborated in detail, and this model endured in Europe
for 15 centuries.  But it was another Greek natural philosopher,
Hipparchus, who made one of the first applications of mathematical
principles that we now consider to be in the realm of statistics.
Finding scatter in Bablylonian measurements of the length of a
year, defined as the time between solstices, he took the middle of
the range -- rather than the mean or median -- for the best value.

This is but one of many discussions of statistical issues in the
history of astronomy.  Ptolemy estimated parameters of a
non-linear cosmological model using a minimax goodness-of-fit
method.  Al-Biruni discussed the dangers of propagating errors
from inaccurate instruments and inattentive observers. While some
Medieval scholars advised against the acquisition of repeated
measurements, fearing that errors would compound rather than
compensate for each other, the usefulnes of the mean to increase
precision was demonstrated with great success by Tycho Brahe.

During the 19th century, several elements of modern mathematical
statistics were developed in the context of celestial mechanics,
where the application of Newtonian theory to solar system
phenomena gave astonishingly precise and self-consistent
quantitative inferences. Legendre developed $L_2$ least squares
parameter estimation to model cometary orbits. The least-squares
method became an instant success in European astronomy and
geodesy.  Other astronomers and physicists contributed to
statistics: Huygens wrote a book on probability in games of
chance; Newton developed an interpolation procedure; Halley laid
foundations of actuarial science; Quetelet worked on statistical
approaches to social sciences; Bessel first used the concept of
"probable error"; and Airy wrote a volume on the theory of errors.

But the two fields diverged in the late-19th and 20th centuries.
Astronomy leaped onto the advances of physics -- electromagnetism,
thermodynamics, quantum mechanics and general relativity -- to
understand the physical nature of stars, galaxies and the Universe
as a whole.  A subfield called ``statistical astronomy'' was still
present but concentrated on rather narrow issues involving star
counts and Galactic structure \cite{Trumpler53}. Statistics
concentrated on analytical approaches. It found its principle
applications in social sciences, biometrical sciences and in
practical industries ({\it e.g.}, Sir R.\ A.\ Fisher's employment
by the British agricultural service).

\section{Statistical needs of astronomy today}

Contemporary astronomy abounds in questions of a statistical
nature.  In addition to exploratory data analysis and simple
heuristic (usually linear) modeling common in other fields,
astronomers also often interpret data in terms of complicated
non-linear models based on deterministic astrophysical processes.
The phenomena studied must obey known behaviors of atomic and
nuclear physics, gravitation and mechanics, thermodynamics and
radiative processes, and so forth. `Modeling' data may thus
involves both the selection of a model family based on an
astrophysical understanding of the conditions under study, and a
statistical effort to find parameters for the specified model.  A
wide variety of issues thus arise:
\begin{itemize}

\item Does an observed group of stars (or galaxies or molecular
clouds or $\gamma$-ray sources) constitute a typical and unbiased
sample of the vast underlying population of similar objects?

\item When and how should we divide/classify these objects into 2,
3 or more subclasses?

\item What is the intrinsic physical relationship between two or
more properties of a class of objects, especially when confounding
variables or observational selection effects are present?

\item How do we answer such questions in the presence of
observations with measurements errors and flux limits?

\item When is a blip in a spectrum (or image or time series) a
real signal rather than a random event from Gaussian (or often
Poissonian) noise or confounding variables?

\item How do we interpret the vast range of temporally variable
objects: periodic signals from rotating stars or orbiting
extrasolar planets, stochastic signals from accreting neutron
stars or black holes, explosive signals from magnetic reconnection
flares or $\gamma$-ray bursts?

\item How do we model the points in 2, 3, ..., 6-dimensional
points representing photons in an image, galaxies in the Universe,
Galactic stars in phase space?

\item How do we quantify continuous structures seen in the sky
such as the cosmic microwave background, the interstellar and
intergalactic gaseous media?

\item How do we fit astronomical spectra to highly non-linear
astrophysical models based on atomic physics and radiative
processes, including confidence limits on the best-fit parameters?

\end{itemize}

From a superficial examination of the astronomical
literature\footnote{Such bibliometric measures are easily
accomplished as the entire astronomical research literature is
on-line (in full text at subscribing institutions) through the
NASA-supported {\it Astrophysics Data System},
http://adsabs.harvard.edu/abstract\_service.html.}, we can show
that such questions are very common today.  Of $\simeq 15,000$
refereed papers published annually, 1\% have ``statistics'' or
``statistical'' in their title, 5\% have ``statistics' in their
abstract, 10\% treat time-variable objects, $5-10$\% (est.)
present or analyze multivariate datasets, and $5-10$\% (est.) fit
parametric models.  Accounting for overlaps, we roughly estimate
that around $\simeq 3,000$ distinct studies each year require
non-trivial statistical methodologies. Roughly 10\% of these are
principally involved with statistical methods; indeed, some of
these purport to develop new methods or improve on established
ones.

\section{Astrostatistics today}

We thus find that astronomy and astrophysics today requires a vast
range of statistical capabilities.  In statistical jargon, it
helps for astronomers to know something about: sampling theory,
survival analysis with censoring and truncation, measurement error
models, multivariate classification and analysis, harmonic and
autoregressive time series analysis, wavelet analysis, spatial
point processes and continuous surfaces, density estimation,
linear and non-linear regression, model selection, and bootstrap
resampling. In some cases, astronomers need combinations of
methodologies that have not yet been fully developed (\S 7 below).

Faced with such a complex of challenges, mechanical exposure to a
wider variety of techniques is a necessary but not sufficient
prerequisite for high-quality statistical analyses.  Astronomers
also need to be imbued with established principles of statistical
inference; {\it e.g.}, hypothesis testing and parameter
estimation, nonparametric and parametric inference, Bayesian and
frequentist approaches, and the assumptions underlying and
applicability conditions for any given statistical method.

Unfortunately, we find that the majority of the thousands of
astronomical studies requiring statistical analyses use a very
limited set of classical methods.  The most common tools used by
astronomers are: Fourier transforms for temporal analysis
(developed by Fourier in 1807), least squares regression and
$\chi^2$ goodness-of-fit (Legendre in 1805, Pearson in 1900,
Fisher in 1924), the nonparametric Kolmogorov-Smirnov 1- and
2-sample nonparametric tests (Kolmogorov in 1933), and principal
components analysis for multivariate tables (Hotelling in 1936).

Even traditional methods are often misused.  Feigelson \& Babu [9]
found that astronomers use interchangeably up to 6 different fits
for bivariate linear least squares regression: ordinary least
squares (OLS), inverse regression, orthogonal regression, major
axis regression, the OLS mean, and the OLS bisector.  Not only did
this lead to confusion in comparing studies ({\it e.g.}, in
measuring the expansion of the Universe via Hubble's constant,
$H_o$), but astronomers did not realize that the confidence
intervals on the fitted parameters can not be correctly estimated
with standard analytical formulae. Similarly, Protassov et al.\
[24] found that the majority of astronomical applications of the
$F$ test, or more generally the likelihood ratio test, are
inconsistent with asymptotic statistical theory.

But, while the {\it average} astronomical study is limited to
often-improper usage of a limited repertoire of statistical
methods, a significant {\it tail of outliers} are much more
sophisticated.  The maximization of likelihoods, often developed
specially for the problem at hand, is perhaps the most common of
these improvements.  Bayesian approaches are also becoming
increasingly in vogue.

In a number of cases, sometimes buried in technical appendices of
observational papers, astronomers independently develop
statistical methods. Some of these are rediscoveries of known
procedures; for example, Avni et al.\ [2] and others recovered
elements of survival analysis for treatments of left-censored data
arising from nondetections of known objects. Some are quite
possibly mathematically incorrect; such as various revisions to
$\chi^2$ for Poissonian data that assume the resulting statistic
still follows the $\chi^2$ distribution.  On rare occasions, truly
new and correct methods have emerged;  for example, astrophysicist
Lynden-Bell [19] discovered the maximum-likelihood estimator for a
randomly truncated dataset, for which the theoretical validity was
later established by statistician Woodroofe [31].

A growing group of astronomers, recognizing the potential for new
liaisons with the accomplishments of modern statistics, have
promoted astrostatistical innovation through cross-disciplinary
meetings and collaborations.  Fionn Murtagh, an applied
mathematician at Queen's University (Belfast) with long experience
in astronomy, and his colleagues have run conferences and authored
many useful monographs ({\it e.g.}, [16], [17], [22] and [27]). We
at Penn State have run a series of {\it Statistical Challenges in
Modern Astronomy} meetings with both communities in attendance
({\it e.g.}, [3] and [10]). Alanna Connors has organized brief
statistics sessions at large astronomy meetings, and we have
organized brief astronomy sessions at large {\it Joint Statistical
meetings}. We wrote a short volume called {\it Astrostatistics}
[3] intended to familiarize scholars in one discipline with
relevant issues in the other discipline.  Other series conferences
are devoted to technical issues in astronomical data analysis but
typically have limited participation by statisticians.  These
include the dozen {\it Astronomical Data Analysis Software and
Systems} ({\it e.g.}, [23]), several Erice workshops on {\it Data
Analysis in Astronomy} ({\it e.g.}, [8]), and the new SPIE {\it
Astronomical Data Analysis} conferences ({\it e.g.}, [26]).

Most importantly, several powerful astrostatistical research
collaborations have emerged.  At Harvard University and the
Smithsonian Astrophysical Observatory, David van Dyk worked with
scientists at the $Chandra$\footnote{The $Chandra$ X-ray
Observatory is one of NASA's Great Observatories.  It was launched
in 1999 with a total budget around \$2 billion.}  X-ray Center on
several issues, particularly Bayesian approaches to parametric
modeling of spectra in light of complicated instrumental effects.
At Carnegie Mellon University and the University of Pittsburgh,
the Pittsburgh Computational Astrophysics group addressed several
issues, such as developing powerful techniques for multivariate
classification of extremely large datasets and applying
nonparametric regression methods to cosmology.  Both of these
groups involved academics, researchers and graduate students from
both fields working closely for several years to achieve a
critical mass of cross-disciplinary capabilities.

Other astrostatistical collaborations must be mentioned. David
Donoho (Statistics at Stanford University) works with Jeffrey
Scargle (NASA Ames Research Center) and others on applying
advanced wavelet methods to astronomical problems.  James Berger
(Statistics at Duke University) has worked with astronomers
William Jefferys (University of Texas), Thomas Loredo (Cornell
University), and Alanna Connors (Eureka Inc.) on Bayesian
methodologies for astronomy.  Bradley Efron (Statistics at
Stanford University) has worked with astrophysicist Veh\'e
Petrosian (also at Stanford) on survival methods for interpreting
$\gamma$-ray bursts.  Philip Stark (Statistics at University of
California, Berkeley) has collaborated with solar physicists in
the $GONG$ program to improve analysis of oscillations of the Sun
(helioseismology).  More such collaborations exist in the U.S.,
Europe and elsewhere.

\section{The Virtual Observatory: A new imperative for astrostatistics}

A major new trend is emerging in observational astronomy with the
production of huge, uniform, multivariate databases from
specialized survey projects and telescopes\footnote{An enormous
collection of catalogs, and some of the underlying imaging and
spectral databases, are already available on-line. Access to many
catalogs is provided by Vizier (http://vizier.u-strasbg.fr). The
NASA Extragalactic Database (NED, http://ned.ipac.caltech.edu),
SIMBAD stellar database (http://simbad.u-strasbg.fr), and ADS
(footnote 2) give integrated access to many catalogs and
bibliographic information. Raw data are available from all U.S.\
space-based observatories; see, for example, the Multi-mission
Archive at Space Telescope (MAST, http://archive.stsci.edu) and
High Energy Astrophysics Science Archive Research Center (HEASARC,
http://heasarc.gsfc.nasa.gov).}.  But they are heterogeneous in
character, reside at widely dispersed locations, and accessed
through different database systems.  Examples include:
\begin{enumerate}

\item $10^8-10^9$-object catalogs of stars and stellar
extragalactic objects ({\it i.e.}, quasars).  These include the
all-sky photographic optical USNO-B1 catalog, the all-sky
near-infrared 2MASS catalog, and the wide-field Sloan Digital Sky
Survey (SDSS). Five to ten photometric values, each with measured
heteroscedastic measurement errors ({\it i.e.}, different for each
data point), are available for each object.

\item $10^5-10^6$-galaxy redshift catalogs from the 2-degree Field
(2dF) and SDSS spectroscopic surveys.  The main goal is
characterization of the hierarchical, nonlinear and anisotropic
clustering of galaxies in a 3-dimensional space. But the datasets
also include spectra for each galaxy each with $10^3$ independent
measurements.

\item $10^5-10^6$-source catalogs from various multiwavelength
wide-field surveys such as the NRAO Very Large Array Sky Survey in
one radio band, the InfraRed Astronomical Satellite Faint Source
catalog in four infrared bands, the Hipparcos and Tycho catalogs
of star distances and motions, and the X-ray Multimirror Mission
Serendipitous Source Catalogue in several X-ray bands now in
progress.  These catalogs are typically accompanied by large image
libraries.

\item $10^2-10^4$-object samples of well-characterized pre-main
sequence stars, binary stars, variable stars, pulsars,
interstellar clouds and nebulae, nearby galaxies, active galactic
nuclei, gamma-ray bursts and so forth.  There are dozens of such
samples with typically $10-20$ catalogued properties and often
with accompanying 1-, 2- or 3-dimensional images or spectra.

\item Perhaps the most ambitious of such surveys is the planned
Large-aperture Synoptic Survey Telescope (LSST) which will survey
much of the entire optical sky every few nights.  It is expected
to generate raw databases in excess of 10 PBy (petabyte) and
catalogs with $10^{10}$ entries.

\end{enumerate}

An international effort known as the Virtual Observatory (VO) is
now underway to coordinate and federate these diverse databases,
making them readily accessible to the scientific user
\citep{Szalay01, Brunner01}. Considerable progress is being made
in the establishment of the necessary data and metadata
infrastructure and standards, interoperability issues, data
mining, and technology demonstration prototype
services\footnote{See http://www.ivoa.net and /http://us-vo.org
for entry into Virtual Observatory projects.}. But scientific
discovery requires more than effective recovery and distribution
of information.  After the astronomer obtains the data of
interest, tools are needed to explore the datasets. How do we
identify correlations and anomalies within the datasets? How do we
classify the sources to isolate subpopulations of astrophysical
interest? How do we use the data to constrain astrophysical
interpretation, which often involve highly non-linear parametric
functions derived from fields such as physical cosmology, stellar
structure or atomic physics?  These questions lie under the aegis
of statistics.

A particular problem relevant to statistical computing is that,
while the speed of CPUs and the capacity of inexpensive hard disks
rise rapidly, computer memory capacities grow at a slower pace.
Combining the largest optical/near-infrared object catalogs today
produces a table with $>1$ billion objects and up to a dozen
columns of photometric data. Such large datasets effectively
preclude use of all standard multivariate statistical packages and
visualization tools ({\it e.g.}, R and GGobi) which are generally
designed to place the entire database into computer memory.  Even
sorting the data to produce quantiles may be computational
infeasible.

The Virtual Observatory of the 21st century thus presents new
challenges to statistical capability in two ways.  First, some new
methodological developments are needed (\S \ref{challenges.sec}).
Second, efficient access to both new and well-established
statistical methods are needed.  No single existing software
package can provide the vast range of needed methods.  We are now
involved in developing a prototype system called {\it VOStat} to
provide statistical capabilities to the VO astronomer. It is based
on concepts of Web services and distributed Grid computing. Here,
the statistical software and computational resources, as well as
the underlying empirical databases, may have heterogeneous
structures and can reside at distant locations.

\section{Some grand methodological challenges for the coming
decade \label{challenges.sec}}

While it is risky to prognosticate the directions of future
research, and judgments will always differ regarding the relative
importance of research goals, we can outline a few ``grand
challenges'' for astrostatistical research for the next decade or
two.

\subsection{Multivariate analysis with measurement errors and
censoring}

Traditional multivariate analysis is designed mainly for
applications in the social and human sciences where the sources of
variance are largely unknowable.  Measurement errors are usually
ignored, or are considered to be exogenous variables in the
parametric models \cite{Fuller87}.  But astrophysicists often
devote as much effort to precise determination of their errors as
they devote to the measurements of the quantities of interest. The
instruments are carefully calibrated to reduce systematic
uncertainties, and background levels and random fluctuations are
carefully evaluated to determine random errors. Except in the
simple case of bivariate regression \citep{Boggs87, Feigelson92,
Akritas96}, this information on measurement errors is usually
squandered.

While heteroscedastic measurement errors with known variances is
common in all physical sciences, only astronomy frequently has
nondetections when observations are made at new wavelengths of
known objects. These are datapoints where the signal lies below
(say) 3 times the noise level. Here again, modern statistics has
insufficient tools. Survival analysis for censored data assumes
that the value below which the data point must lie is known with
infinite precision, rather than being generated from a
distribution of noise. Astronomer Herman Marshall [20] makes an
interesting attempt to synthesize measurement errors and
nondetections, but statistician Leon Gleser [14] argues that he
has only recovered Fisher's failed theory of fiducial
distributions. Addressing this issue in a self-consistent
statistical theory is a profound challenges that lies at the heart
of interpreting the data astronomers obtain at the telescope.

\subsection{Statistical inference and visualization with
very-large-N datasets}

The need for computational software for extremely large databases
-- multi-terabyte image and spectrum libraries and multi-billion
object catalogs -- is discussed in section 4. A suite of
approximate methods based on flowing data streams or adaptive
sampling of large datasets resident on hard disks should be
sought.  Visualization methods involving smoothing,
multidimensional shading and variable transparency, should be
brought into the astronomer's toolbox.  Here, considerable work is
being conducted by computer scientists and applied mathematicians
in other applied fields so that independent development by
astrostatisticians  might not be necessary to achieve certain
goals.

\subsection{A cookbook for construction of likelihoods and Bayesian
computation}

While the concepts of likelihoods and their applications in
maximum likelihood estimation, Bayes Theorem and Bayes factors are
becoming increasingly well-known in astronomical research, the
applications to real-life problems is still an art for the expert
rather than a tool for the masses. Part of the problem is
conceptual; astronomers need training in how to construct
likelihoods for familiar parametric situations ({\it e.g.}, power
law distributions or a Poisson process). Part of the problem is
computational; astronomers need methods and software for the
oft-complex computations.  Many such methods, such as Markov chain
Monte Carlo, are already well-established and can be directly
adopted for astronomy [13]. For example, astronomers are often not
fully aware of the broad applicability of the EM Algorithm for
maximizing likelihoods \cite{McLachlan97}\footnote{The seminal
study of the EM Algorithm is Dempster, Laird \& Rubin in 1977 [7],
which is one of the most frequently cited papers in statistics.
However, the method was independently derived three years earlier
by astronomer Leon Lucy [18] as an ``iterative technique for the
rectification of observed distributions'' based on Bayes' Theorem.
This study is widely cited in the astronomical literature; its
most frequent application is in image deconvolution where it is
known as the Lucy-Richardson algorithm.}.

\subsection{Links between astrophysical theory and wavelets}

Wavelet analysis has become a powerful and sophisticated tool for
the study of features in data.  Originally intended mainly for
modelling time series, astronomers also use it increasingly for
spatial analysis of images \cite{Freeman02, Scargle98}.  In some
ways it can be viewed as a generalization of Fourier analysis in
which the basis function need not be sinusoidal in shape and, most
importantly, the pattern need not extend over the entire dataset.
Wavelets are thus effective in quantitatively describing
complicated overlapping structures on many scales, and can also be
used for signal denoising and compression. In addition, wavelets
have a strong mathematical foundation.

Despite its increasing popularity in astronomical applications,
wavelet analysis suffers a profound limitation in comparison with
Fourier analysis.  A peak in a Fourier spectrum is immediately
interpretable as a vibrational, rotational or orbital rotation of
solid bodies.  A bump or a continuum slope in a wavelet
decomposition often has no analogous physically intuitive
interpretation.  We therefore recommend that astrophysicists seek
links between physical theory -- often involving continuous media
such as turbulent plasmas in the interstellar medium and
hierarchical structure formation in the early Universe -- and
wavelets.  One fascinating example is the demonstration that the
wavelet spectrum and Lyapunov exponent of the quasi-periodic X-ray
emission from Sco X-1, which reflects the processes in an
accretion disk around a neutron star, exhibit a transient chaotic
behavior similar to that of water condensing and dripping onto an
automobile windshield or a dripping handrail \citep{Young96}.

\subsection{Time series models for astrophysical phenomena}

The quasi-periodic oscillation of Sco X-1 is only one of many
examples of complex accretional behavior onto neutron stars and
black holes seen in X-ray and $\gamma$-ray astronomy.  The
accreting Galactic black hole GRS 1915+105 exhibits a bewildering
variety of distinct states of stochastic, quasi-periodic and
explosive behaviors.  The prompt emission from gamma-ray bursts
show a fantastic diversity of temporal behaviors from simple
smooth fast-rise-exponential-decays to stochastic spiky profiles.
Violent magnetic reconnection flares on the surfaces of the Sun
and other magnetically active stars also show complex behaviors.
Many of these datasets are multivariate with time series available
in several spectral bands often showing lags or hardness ratio
variations of astrophysical interest.

There are also important astronomical endeavors which seek
astrophysically interesting signals amidst the oft-complex noise
characteristics of the detectors.  The Arecibo, Parkes and VLA
radio telescopes, for example, conduct searches for new radio
pulsars or for extraterrestrial intelligences in nearby planetary
systems.  The Laser Interferometer Gravitational-Wave Observatory
(LIGO) and related detectors search for both continuing periodic
signals and brief bursts from perturbations in space-time
predicted by Einstein's General Relativity.  Here the signals
sought are orders of magnitude fainter than instrumental
variations.

\section{Infrastructure needed to advance astrostatistics}

The current quality of statistical analyses in astronomical
research often begs for improvment.  There is both inadequate
research on important new challenges (\S \ref{challenges.sec}) and
inadequate application of known advanced methods to astronomical
problems (\S 3).  Astronomy clearly needs needs a strong and rapid
surge of energy in statistical expertise.  Three types of
activities should be promoted: \begin{description}

\item[Cross-training] ~~~ In the U.S., the typical curriculum
leading to a career in astronomical research requires zero or one
course in statistics at the undergraduate level, and zero at the
graduate level. Analogously, the curriculum of statisticians
includes virtually no coursework in astronomy or other physical
science. While statisticians can learn basics from ``Astronomy
101'' courses given at all universities, the statistical training
of astronomers is not as easily accomplished. New curricular
products summarizing the applicable statistical subfields, short
training workshops for graduate students and young scientists, and
effective statistical consulting are all needed.

\item[Increased collaborative research] ~~~  While several
astrostatistical research groups are making exciting progress (\S
3), the total effort is too small to impact the bulk of
astronomical research.  Very roughly, astrostatistical funding is
currently \$1M of the \$1B spent annually on astronomical
research.  This fraction is far below that spent in biomedical or
other non-physical-science fields.  Though top academic leaders of
statistics have expressed great enthusiasm for astronomy and
astrostatistics, we can not pull them away from biostatistics and
business applications without a major increase in funding. We
might seek, for example, $10-20$ cross-disciplinary research
groups active at any one time at the end of a decade's growth.

\item[Statistical software] ~~~ For various policy and cultural
reasons, astronomers rarely purchase the large commercial
statistical software packages, preferring to write their own
software as needs arise.  This approach has contributed to the
narrow methodological scope of astronomical research.  Avenues for
improving this situation are emerging.  {\it R} is a large
statistical software package with the flexible command-line
interface preferred by astronomers that has recently emerged
(http://www.r-project.org).  A wide variety of specialized
packages and codes are also available on-line
(http://www.astro.psu.edu/statcodes).  The new Web services
concept being developed within the context of a Virtual
Observatory permits coordinated access to heterogeneous software
developed specifically for astronomical applications.

\end{description}

At Penn State, we are in the early stages of developing a Center
for Astrostatistics to help attain these goals
(http://www.astrostatistics.psu.edu). This is an
inter-disciplinary Center to serve the astronomy and statistics
communities around the nation and worldwide, seeking to bring
advances in statistics into the toolbox of astronomy and
astrophysics. The Center's Web site will maintain the popular {\it
StatCodes}, build an instructional library of $R$ programs,
coordinate with the nascent {\it VOStat} Web service, and develop
an archive of annotated links to selected statistical literature
applicable to astronomy (and vice versa).  The site is also
planned to include tutorial handbooks and curricular products
developed specifically for astrostatistical needs.

\begin{acknowledgments}
We thank the referee, Peter Shawhan (Caltech), for very helpful
improvements to the paper.  This work was supported by NSF grant
DMS-0101360.
\end{acknowledgments}


\begin{thebibliography}{99}

\bibitem{Akritas96} Akritas, M.\ G.\ \& Bershady, M.\ A.\ 1996,
Astrophys.\ J.\ 470, 709

\bibitem{Avni80} Avni, Y., Soltan, A., Tananbaum, H., \& Zamorani,
G.\ 1980, Astrophys.\ J.\ 238, 800

\bibitem{Babu96} Babu, G.\ J.\ \& Feigelson, E.\ D.\ 1996,
Astrostatistics, London, Chapman \& Hall

\bibitem{Babu97} G.\ J.\ Babu \& Feigelson, E.\ D.\ 1997,
Statistical Challenges in Modern Astronomy II, New York, Springer

\bibitem{Boggs87} Boggs, P.\ T., Byrd, R.\ H.\ \& Schnabel, R.\
B.\ 1987, SIAM J.\ Sci.\ Statist.\ Comput.\ 8, 1052

\bibitem{Brunner01} Brunner, R.\ J., Djorgovski, S.\ G. \& Szalay,
A.\ S. (eds.) 2001, Virtual Observatories of the Future, San
Francisco, Astron.\ Soc.\ Pacific

\bibitem{Dempster77} Dempster, A.\ P., Laird, N.\ M.\ \& Rubin,
D.\ B.\ 1977, J.\ Roy.\ Statist.\ Soc.\ Ser.\ B, 39, 1

\bibitem{DiGesu} Di Gesu, V., Duff, M.~J.~B., Heck, A., Maccarone,
M.~C., Scarsi, L., \& Zimmerman, H.~U.\ 1997, Data Analysis in
Astronomy IV,

\bibitem{Feigelson92} Feigelson, E.\ D.\ \& Babu, G.\ J.\ 1992,
Astrophys.\ J.\ 397, 55

\bibitem{Feigelson03} Feigelson, E.\ D.\ \& Babu, G.\ J.\ 2003,
Statistical Challenges in Modern Astronomy, New York, Springer

\bibitem{Freeman02} Freeman, P.~E., Kashyap, V., Rosner, R., \&
Lamb, D.~Q.\ 2002, Astrophys.\ J.\ Suppl., 138, 185

\bibitem{Fuller87} Fuller, W.\ A.\ 1987, Measurement Error Models,
New York, Wiley

\bibitem{Gelman95} Gelman, A., Carlin, J.\ B., Stern, H.\ S.\ \&
Rubin, D.\ B.\ 1995, Bayesian Data Analysis, London, Chapman \&
Hall

\bibitem{Gleser92} Gleser, L.\ J.\ 1992, in Statistical
Challenges in Modern Astronomy (E.\ D.\ Feigelson \& G.\ J.\ Babu,
eds.), New York, Springer, 263

\bibitem{Hald90} Hald, A.\ 1990, A History of Probability and
Statistics and Their Applications before 1750, New York, Wiley

\bibitem{Heck93} Heck, A.~\& Murtagh, F.\ 1993,  Intelligent
Information Retrieval: The Case of Astronomy and Related Space
Sciences, Dordrecht, Kluwer

\bibitem{Jaschek} Jaschek, C.~\& Murtagh, F.\ 1990, Errors,
Bias and Uncertainties in Astronomy, Cambridge, Cambridge Univ.\
Press

\bibitem{Lucy74} Lucy, L.\ B.\ 1974, Astron.\ J.\ 79, 745

\bibitem{LyndenBell71} Lynden-Bell, D.\ 1971, Mon.\ Not.\ Roy.\
Astro.\ Soc.\ 155, 95

\bibitem{Marshall92} Marshall, H.\ L 1992, in Statistical
Challenges in Modern Astronomy (E.\ D.\ Feigelson \& G.\ J.\ Babu,
eds.), New York, Springer, 247

\bibitem{McLachlan97} McLachlan, G.\ J.\ \& Krishnan, T.\ 1997,
The EM Algorithm and Extensions, New York, Wiley

\bibitem{Murtagh87} Murtagh, F. \& Heck, A.\ 1987, Multivariate
Data Analysis, Dordrecht, Reidel

\bibitem{Payne03} Payne, H.~E., Jedrzejewski, R.~I., \& Hook, R.~N.\
2003, Astronomical Data Analysis Software and Systems, San
Francisco, Astro.\ Soc.\ Pacific

\bibitem{Protassov02} Protassov, R., van Dyk, D.~A., Connors, A.,
Kashyap, V.~L., \& Siemiginowska, A.\ 2002, Astrophys.\ J., 571,
545

\bibitem{Scargle98} Scargle, J.~D.\ 1998, Astrophys.\ J., 504, 405

\bibitem{Starck02} Starck, J.\-L.\ \& Murtagh, F.\ (eds.) 2002,
Astronomical Data Analysis II, Proc. SPIE, vol 4847

\bibitem{Starck03} Starck, J.\-L.\ \& Murtagh, F.\ 2003,
Astronomical Image and Data Analysis, New York, Springer

\bibitem{Stigler86} Stigler, S.\ M.\ 1986, {\it The History of Statistics: The
Measurement of Uncertainty before 1900}, Cambridge, Harvard Univ.\
Press

\bibitem{Szalay01} Szalay, A.\ \& Gray, J.\ 2001, Science, 293,
2037

\bibitem{Trumpler53} Trumpler, R.\ J.\ \& Weaver, H.\ A., 1953, {\it Statistical
Astronomy}, Berkeley, Univ.\ of California Press

\bibitem{Woodroofe85} Woodroofe, M.\ 1985, Annals Statist.\ 13,
163

\bibitem{Young96} Young, K.~\& Scargle, J.~D.\ 1996, Astrophys.\ J.,
468, 617


\end{thebibliography}
\end{document}